\documentclass[nofootinbib,showkeys,showpacs,11pt]{revtex4}
\pdfoutput=1
\usepackage{xcolor}
\usepackage{amsmath,amsfonts,amssymb,amsthm}
\usepackage{graphicx}
\usepackage{placeins}
\usepackage{epstopdf}
\usepackage{subfigure}
\usepackage{parskip}
\usepackage[compact]{titlesec}
\usepackage[none]{hyphenat}
\usepackage[hyperfootnotes=false]{hyperref}
\hypersetup{colorlinks=true,citecolor=blue,linkcolor=blue,urlcolor=blue}
\newcommand{\ef}{\operatorname{Erf}}

\titlespacing{\section}{0pt}{1em}{0em}
\titlespacing{\subsection}{0pt}{1em}{0em}

\begin{document}

\title{\hfill { Nonlinear Dynamics (2014).\\
\hfill DOI: \href{http://www.dx.doi.org/10.1007/s11071-014-1674-9}{10.1007/s11071-014-1674-9}}\\ \vspace{1 cm} \large Analytical stable Gaussian soliton supported by a parity-time-symmetric potential with power-law nonlinearity}



\author{Bikashkali Midya}

\affiliation{\small Physique Nucl\'eare et Physique Quantique, Universite libre de Bruxelles, Brussels, Belgium.\vspace{.6 cm}}  
 \email{bmidya@ulb.ac.be}

\begin{abstract}
We address the existence and stability of spatial localized modes supported by a parity-time-symmetric complex potential in the presence of power-law nonlinearity. The analytical expressions of the localized modes, which are Gaussian in nature, are obtained in both (1+1) and (2+1) dimensions. A linear stability analysis corroborated by the direct numerical simulations reveals that these analytical localized modes can propagate stably for a wide range of the potential parameters and for various order nonlinearities. Some dynamical characteristics of these solutions, such as the power and the transverse power-flow density, are also examined.

\keywords{Parity-time symmetry, Gain and loss, Nonlinear Schr\"odinger equation, Optical soliton \vspace{.5 cm}}
\pacs{42.65.Tg, 42.65.Wi, 42.25.Bs, 11.30.Er}

\end{abstract}
\maketitle

\section{Introduction}\label{intro}
Spatial solitons are the localized bound states that can preserve their shapes during propagation in a nonlinear media. They exist by virtue of the balance between diffraction and nonlinearity. Recently, the propagation of such optical beam in complex nonlinear media, featuring the parity-time (PT)-symmetry, has been a subject of intense investigation \cite{Mu+08,Ma+08,AK12,HZ12,AK11,Ya12,AB12,Mal13,MR13,MR14,Khare12,Gu12,ZK12,Li+11,XD14,DWZ14,Sh11,HH12,Ch+14,DH14,Bl+13}. The PT-symmetry originates from quantum mechanics: A wide class of complex Hamiltonians, although non-Hermitian, can still exhibit real and discrete energy eigenvalues provided that they respect the combined symmetry of parity ($\mathcal{P}$) and time-reversal ($\mathcal{T}$) operators \cite{Be07}. The  operator $\mathcal{P}$, responsible for spatial reflection, is defined through the operations $p \rightarrow -p, x \rightarrow -x$, while the operator $\mathcal{T}$ leads to $p \rightarrow -p, x \rightarrow x$ and to complex conjugation $i \rightarrow -i$. A necessary condition for a Hamiltonian to be PT-symmetric is that the potential function $V_{PT} = V(x) + i W(x)$ should satisfy the condition $V_{PT}(x) = V_{PT}^*(-x)$, with $*$ denoting the complex conjugation. The connection, bridged by the Schr\"odinger equation, between the quantum mechanics and the optical field made it possible to realize the idea of PT-symmetry in optical structures. In optics, the complex refractive index plays the role of a PT-symmetric potential such that $V(x)$, the index guiding, is an even function of position and $W(x)$, the gain-loss profile, is an odd function. Thus, optical structures having PT-symmetry can realistically be implemented through a judicious inclusion of balanced gain and loss in optical waveguides. A series of recent experiments on such optical structures lead to altogether new dynamics such as abrupt phase transition, power oscillation, double refraction, and appearance of stable optical soliton etc \cite{Ru+10,Gu+09,KGM08}.

 Optical soliton supported by a PT-symmetric complex potential may exhibit two parametric regions: a region where such soliton propagates stably whereas the other region corresponds to the unstable propagation. A threshold, therefore, exists between these two regions; below this threshold value. the solitons are stable, whereas they are unstable above the threshold. The existence of various types of stable solitons in diverse complex potentials with gain and loss has been reported. These works include, for example, solitons in nonlinear lattices \cite{AK11,HZ12}, dark soliton and vortices \cite{AK12}, Gray soliton \cite{Li+11}, stable bright soliton in defocusing Kerr media \cite{Sh11} and stable light bullet solutions \cite{DWZ14}. Exact closed form nonlinear localized modes in PT-symmetric optical media with competing gain-loss \cite{MR14} and with competing nonlinearity \cite{Khare12} are reported. The role of PT-symmetry has been investigated in nonlinear tunneling \cite{XD14} and on soliton propagation in nonlinear couplers \cite{DM11,DM12,AB12}. Two dimensional solitons in PT-symmetric periodic \cite{Ya12}, nonlocal \cite{ZBH12} and inhomogeneous \cite{DW14} nonlinear media are also studied.  In most of these works, the localized solutions are obtained with the help of some numerical techniques. In fact, there are very few models which possess exact analytical expressions for the localized modes \cite{Mu+08a,MR13,MR14,Khare12,Mal13,Xu14,DM11a,WDW14}. It is worth mentioning here about some recent analytical approaches for obtaining exact solution of the nonlinear Schr\"odinger equation (NLSE), e.g., similarity transformation method \cite{Zh13}, the Hirota method \cite{LTL13} and Painleve-integrability analysis \cite{LP13}.     
   
In the present paper, we consider a parity-time-symmetric complex potential [given in equation (\ref{e3})], which admits analytical localized solution in the presence of power-law nonlinearities including the Kerr one. The exact expressions of such solution, which are Gaussian in nature, are obtained in both $(1+1)$ and $(2+1)$ dimensional NLSE. A linear stability analysis corroborated by the direct numerical simulations reveals that these analytical localized modes can propagate stably for a wide range of the potential parameters and for higher order nonlinearities. Nature of the total power and the transverse power-flow density associated with these solutions are also discussed. The rest of the paper is organized as follows. In section II, we have briefly described the mathematical model of the optical beam propagation and obtained the corresponding exact solutions in one-dimension (1D). The linear stability of these solutions is discussed in section IIA. Section III deals with the analytical localized modes and their linear stability in two-dimensions (2D). Conclusions are drawn in Section IV.

\section{Exact Gaussian solitons in a 1D PT-symmetric potentials} \label{sec1}
We consider beam propagation along the $z$ direction in a nonlinear medium of non-Kerr index with a complex transverse refractive
index profile. Evolution of the dimensionless
light field envelope $\Psi(x,z)$ in a single PT cell is governed by the $(1+1)$ dimensional NLSE \cite{Ma+08,Mu+08a}:
 \begin{equation}
 i \frac{\partial \Psi}{\partial z} + \frac{\partial^2 \Psi}{\partial x^2} + \left[V(x)+ i W(x)\right] \Psi + \sigma |\Psi|^{2m} \Psi = 0, \label{e1}
\end{equation}
where the transverse $x$ and longitudinal $z$ coordinates are
scaled in the terms of diffraction length and beam width, respectively. $\sigma= \pm 1$ corresponds to the self-focusing and self-defocusing nonlinearities, respectively. Here, $m$ is a positive integer which determines the order of the nonlinearity. For $m=1$ one has the Kerr nonlinearity, for $m=2$ the quintic, for $m=3$ the septic, etc. $V(x)$ and $W(x)$ are the real and imaginary parts of the PT-symmetric complex potential 
which satisfy $V(-x) = V(x)$ and $W(-x) = -W(x)$. In this paper, we choose $V(x)$ and $W(x)$, as
\begin{equation}
V(x) = V_0 x^2 + V_1 e^{-2 a^2 x^2}, ~~~ W(x) = W_0 x e^{-a^2 x^2} \label{e3}
\end{equation}
where the potential parameters $V_0, V_1$, $W_0$ and $a$ are real.  First three parameters determine the depth of the potential, and the last one is responsible for various width. Note here that in the practical application of an optical beam propagation, it is desirable to consider a periodic potential. However, we shall show in the following that a single PT cell characterized by the above potential is capable to produce stable solitonic solutions by suppressing the collapse caused by the diffraction. The stationary solution of the NLSE (\ref{e1}) can be determined by assuming $\Psi(x,z) = \phi(x) e^{i \beta z}$,
where the complex valued function $\phi(x)$ describes, in general, the soliton and $\beta$ is a real propagation constant. 
Substitution of this expression of $\Psi(x,z)$ into the Eq. (\ref{e1}) yields an ordinary nonlinear differential equation of $\phi(x)$ :
\begin{equation}
\frac{d^2 \phi}{d x^2} + \left[V_0 x^2 + V_1 e^{-2 a^2 x^2} + i W_0 x e^{-a^2 x^2}\right] \phi + \sigma |\phi|^{2m} \phi = \beta \phi. \label{e31}
\end{equation}
The above equation (\ref{e31}) can be solved numerically to find the localized solutions $\phi(x)$. However, we find that the aforementioned equation admits localized solutions in simple analytical form: 
\begin{equation}
\phi(x) = \phi_0 ~ e^{- \frac{a^2 x^2}{m} + i \frac{m W_0 \sqrt{\pi}}{4 a^3 (m+2)} \ef(a x)} \label{e32}
\end{equation}
where $\ef(x)$ denotes the error function and
\begin{equation}
V_0 = -\frac{4 a^4}{m^2}, ~ \phi_0 = \left|\frac{1}{\sigma} \left(\frac{m^2 W_0^2}{4 a^4 (m+2)^2}-V_1\right)\right|^{\frac{1}{2m}},~\beta = -\frac{2 a^2}{m}
\end{equation}
and $V_1$, $W_0$ are arbitrary. These solutions, which are Gaussian in nature, are valid for all values of the nonlinearity parameter $m$ and for both the self-focusing and self-defocusing nonlinearities. The real and imaginary parts of this localized solution are shown in figure \ref{fig1}(b) for $m=2, \sigma=1, a= 0.5, V_1 = 3$ and $W_0 = 0.2$. Before doing the linear stability analysis of the solution (\ref{e32}), we look into the following two quantities. The total power of the stationary localized solution is $P = \int_{-\infty}^{\infty} |\phi|^2 dx = \frac{1}{a}\sqrt{\frac{m \pi}{2}} \phi_0^{2}$. Hence, $P$ is always real and positive for both the self-focusing and self-defocusing cases if $a>0$. On the other hand the transverse power or the energy flow-density is given by ${S} = (i/2) [\phi \phi_x^* - \phi^* \phi_x] = \frac{m W_0 \phi_0^{2}}{2 a^2 (m+2)} e^{-\frac{m+2}{m} a^2 x^2}$. Clearly, ${S}$ is everywhere positive if $W_0 > 0$. The positive value of $S$ implies that the energy flow is one-way (from gain toward loss).

\subsection{Linear stability analysis of the exact solutions}
In this section, we analyze the linear stability of the exact localized solutions obtained in previous section. To do this aim, we consider a small perturbation to the localized solution of Eq. (\ref{e1}) in the form 
\begin{equation}
 \Psi(x,z) = \left\{\phi(x)+ [f(x)+g(x)] ~e^{\delta z} + [f^*(x)-g^*(x)]~ e^{\delta^* z}\right\} e^{i \beta z}\label{e30}
\end{equation}
where $f$, $g \ll 1$ are infinitesimal normal mode perturbation eigenfunctions, and $\delta$ is the corresponding eigenvalue. Substituting the perturbed solution into Eq. (\ref{e1}) and linearizing, one obtains the following linear-stability eigenvalue problem \cite{Ya12}:
\begin{equation}
 i \left( \begin{array}{cc}
\mathcal{L}_0 & ~\mathcal{L}_1  \\
 \mathcal{L}_2 & ~-\mathcal{L}_0 \\
\end{array} \right)   ~~~ \left( \begin{array}{c}
f   \\
g \\
 \end{array} \right)=  \delta
 \left( \begin{array}{c}
f \\
g \\
 \end{array} \right)\label{e12}
\end{equation}
where 
\begin{equation}\begin{array}{lll}
\mathcal{L}_0 = \frac{\sigma m}{2} (\phi^2 - \phi^{*2}) |\phi|^{2m-2}\\
\mathcal{L}_1 = \partial_{xx} + (V+i W) + \sigma |\phi|^{2m} + \sigma m |\phi|^{2m-2}\left(|\phi|^2 - \frac{\phi^2 - \phi^{*2}}{2}\right) - \beta\\

\mathcal{L}_2 = \partial_{xx} + (V+i W) + \sigma |\phi|^{2m} + \sigma m |\phi|^{2m-2}\left(|\phi|^2 + \frac{\phi^2 - \phi^{*2}}{2}\right) - \beta
\end{array}
\end{equation}
\begin{figure}[h!]
\includegraphics[width=4.3cm,height=4.75cm]{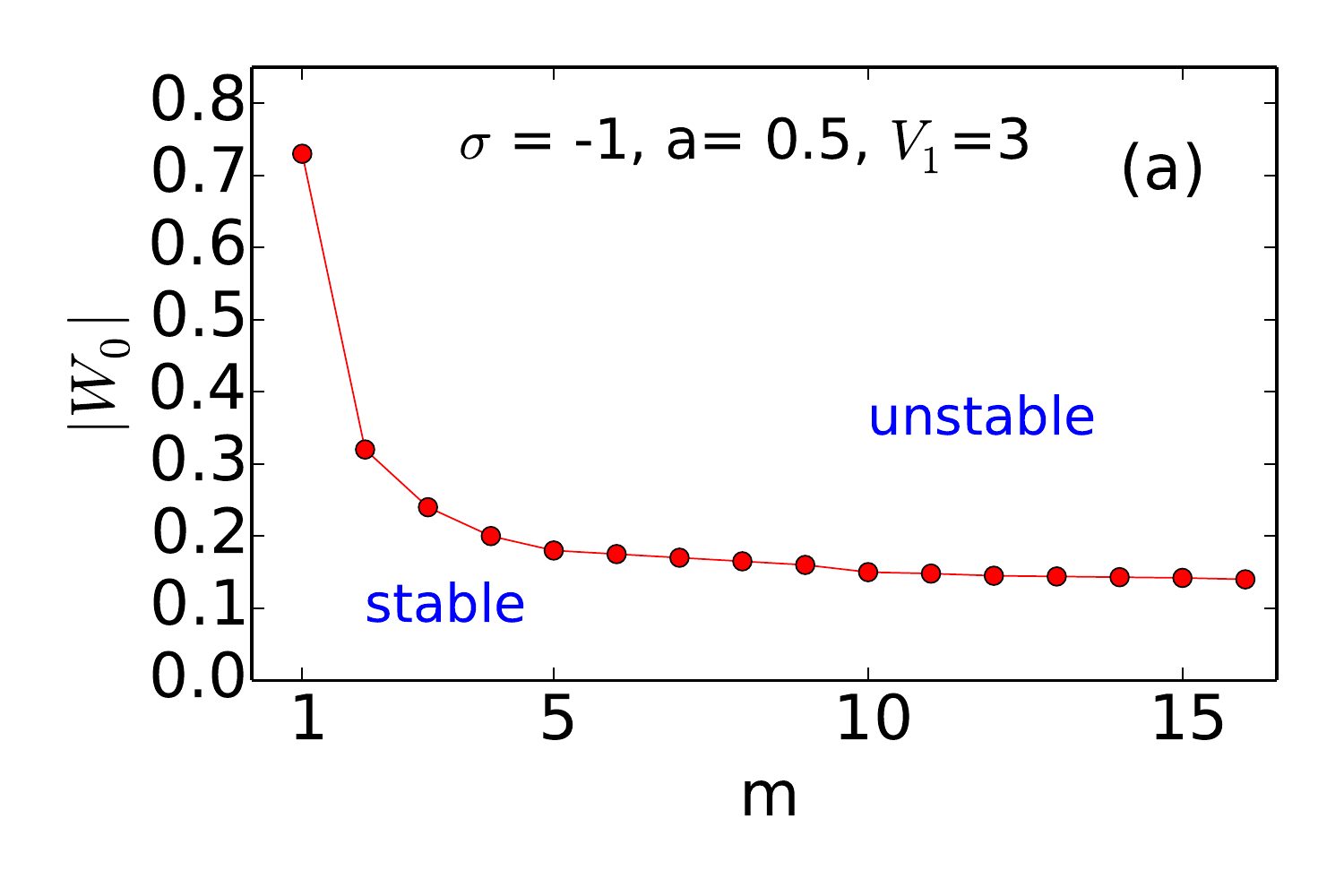} \includegraphics[width=5cm,height=4.5cm]{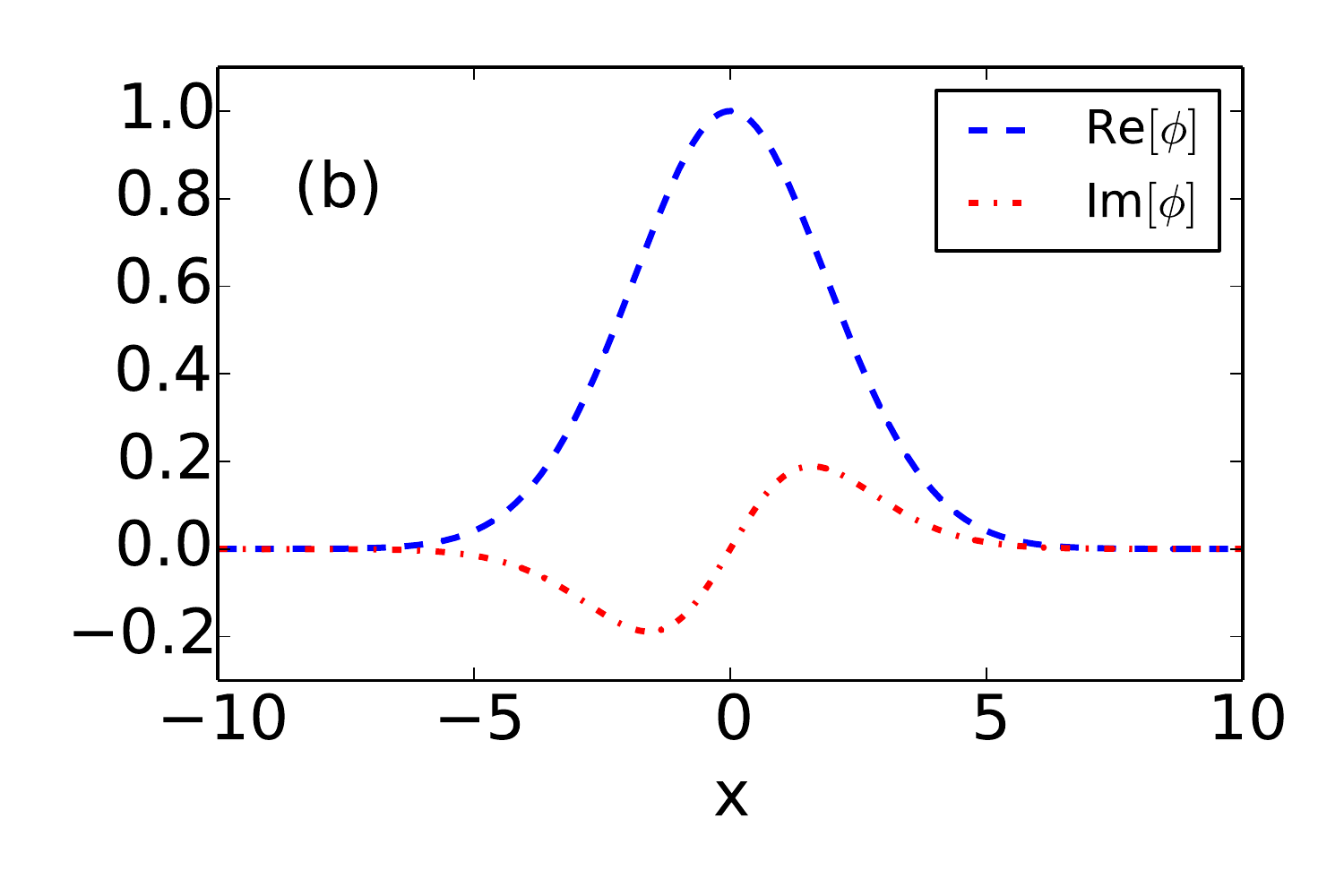}
\includegraphics[width=4.25cm,height=4.5cm]{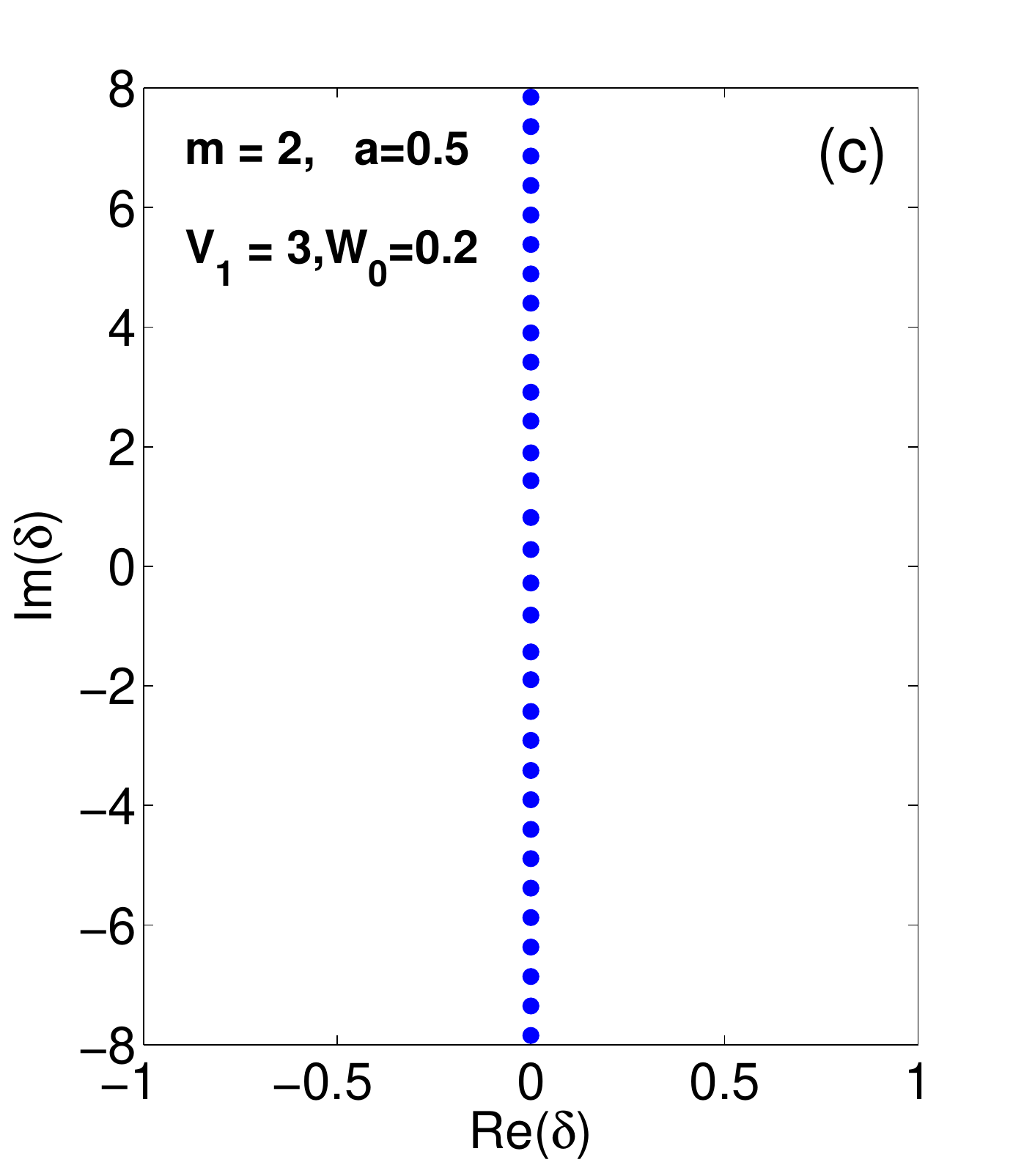} \includegraphics[width=5.25cm,height=4cm]{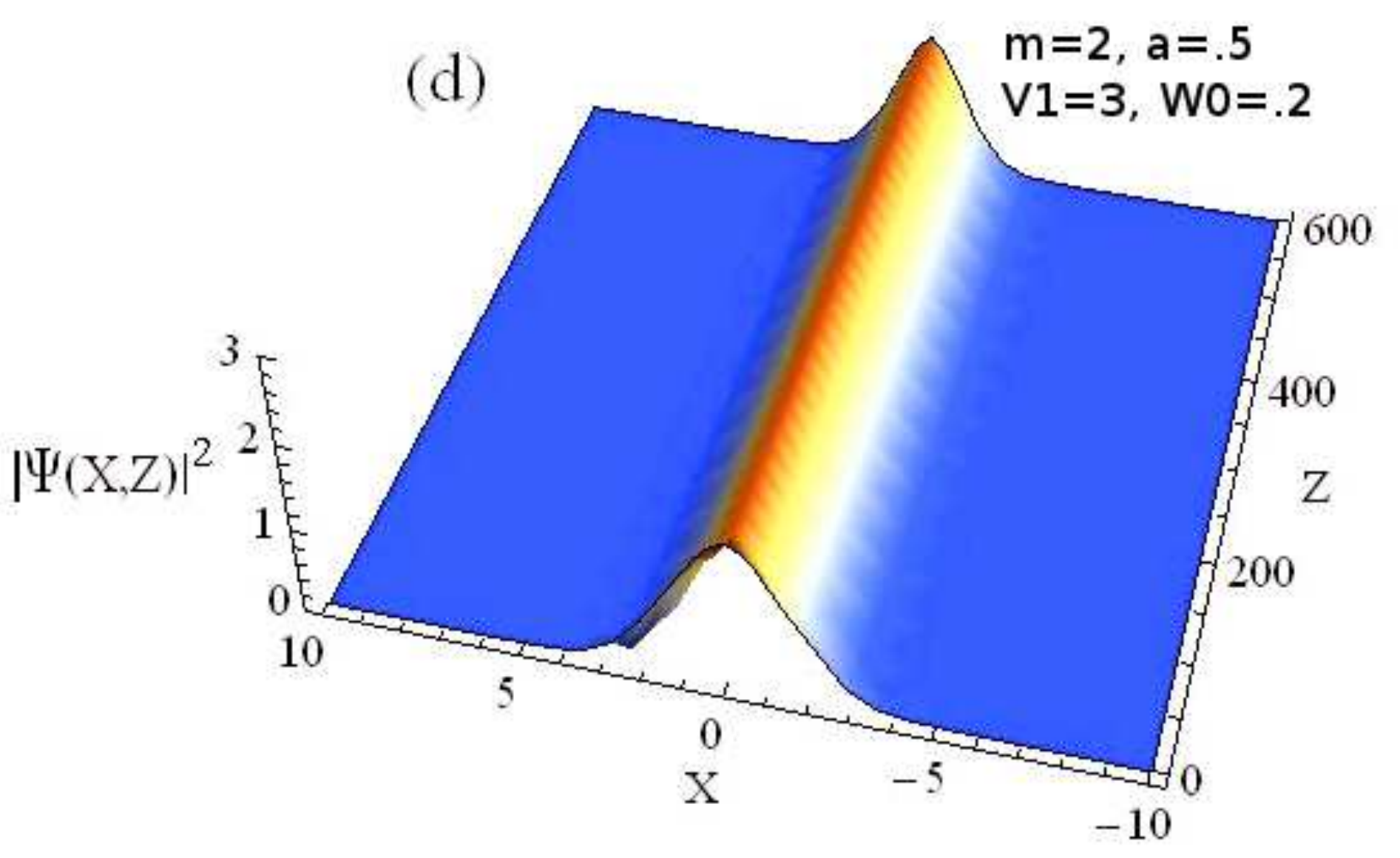}
\includegraphics[width=4cm,height=4.75cm]{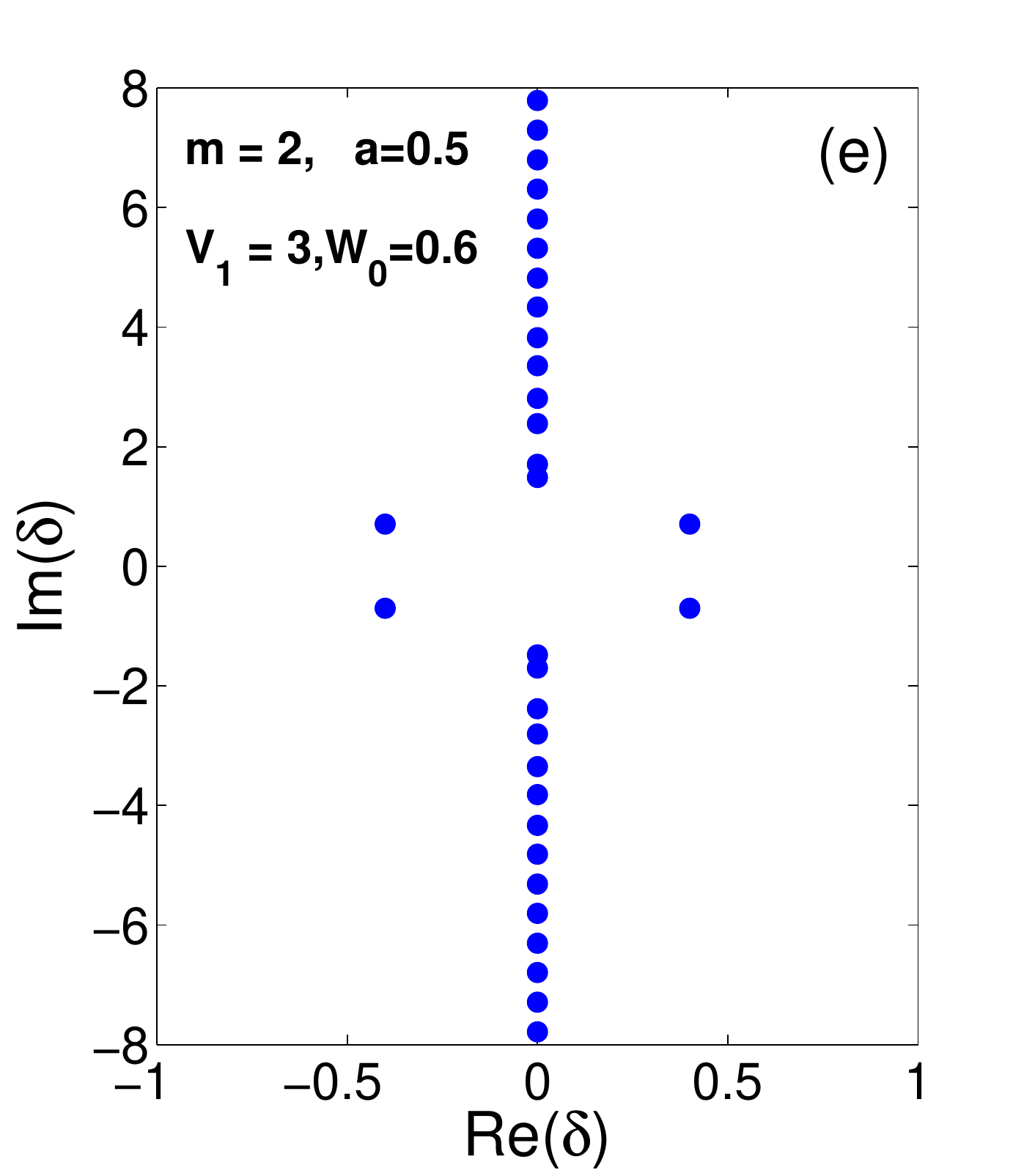} \includegraphics[width=5.25cm,height=4cm]{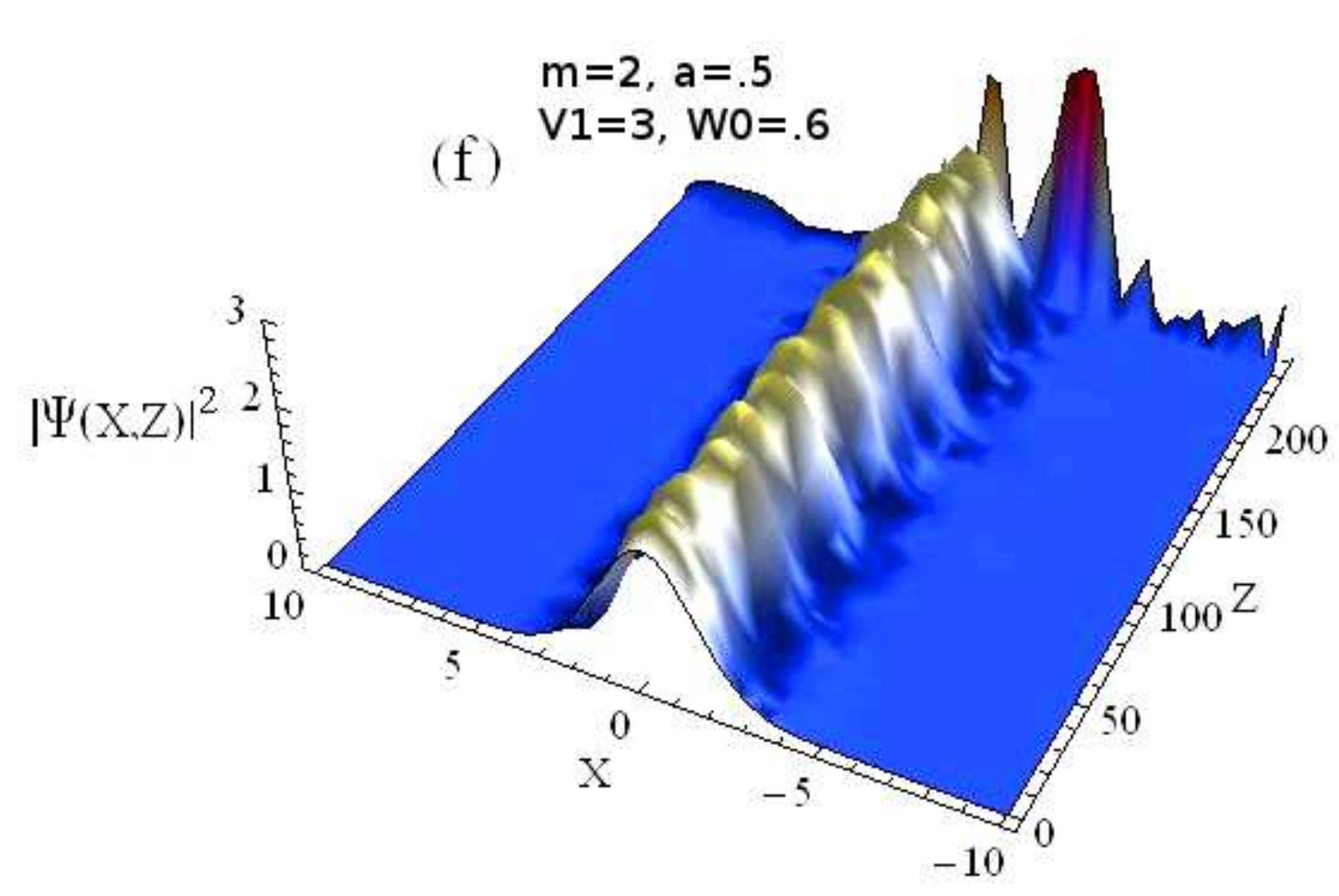}
\caption{(Color online) Plot of the (a) stability region for several values of $m$, (b) the real and imaginary parts of the solution (\ref{e32}) for $m=2$ and for the parameter values $a=0.5, V_1=3, W_0 = 0.2$; (c) and (e) the linear stability eigenvalue spectra; (d) and (f) the stable and unstable intensity evolutions with stability spectra shown in (c) and (e), respectively. In all these cases we have considered self-defocusing nonlinearity $\sigma=-1$.}
\label{fig1}
\end{figure}
\FloatBarrier
The nature of the eigenvalue spectrum of the equation (\ref{e12}) determines the linear stability criterion. Obviously, if there exist a $\delta$ with nonzero positive real part, the solution $\Psi(x,z)$ will grow exponentially upon propagation and thus becomes unstable. On the other hand, in the absence of such $\delta$ the solution becomes linearly stable. Here, we use Fourier collocation method \cite{Ya08} in order to obtain the whole eigenvalue spectra of the above problem (\ref{e12}). 
For numerical computation, here we have restricted $a=0.5$ and  $V_1=3$ and performed the linear stability analysis on the solution (\ref{e32}) for several values of $m$ and for both the self-focusing and self-defocusing nonlinearities.  The results of our analysis for the self-defocusing case are summarized in figure \ref{fig1}(a). Corresponding to a particular $m$, there exists a threshold value of $|W_0|$ below which the localized modes propagate stably and above this threshold value, the modes become unstable. For $m=1$, we find the threshold value as $|W_0| \sim 0.73$, for $m=2$ it is $|W_0| \sim 0.32$ etc. From the figure \ref{fig1}(a), it is clear that the stability region decreases as the order of nonlinearity increases. However, after $m>10$ the threshold value $(|W_0| \sim 0.14)$ is almost same for all $m$. The detailed structure of the soliton, the linear stability spectra and the intensity evolutions for a particular value of $m=2$ are displayed in the remaining figures of fig.\ref{fig1}. The real and imaginary parts of the soliton are plotted in figure \ref{fig1}(b). In figures \ref{fig1}(c) and \ref{fig1}(e), we have shown the linear stability eigenspectra for two different values of $W_0$, one is below the threshold and another is above the threshold, respectively. The former case implies that the localized mode is stable and the latter case implies that it is unstable. To verify the stability analysis results, we have solved the equation (\ref{e1}) by the direct numerical simulation after choosing $\phi(x)$ as the initial profile, i.e., $\Psi(x,0) = \phi(x)$. Corresponding stable and unstable propagations, in agreement with the linear stability analysis, are displayed in figures \ref{fig1}(d) and \ref{fig1}(f), respectively.\\

\begin{figure}[h!]
 \includegraphics[width=4.3cm,height=4.75cm]{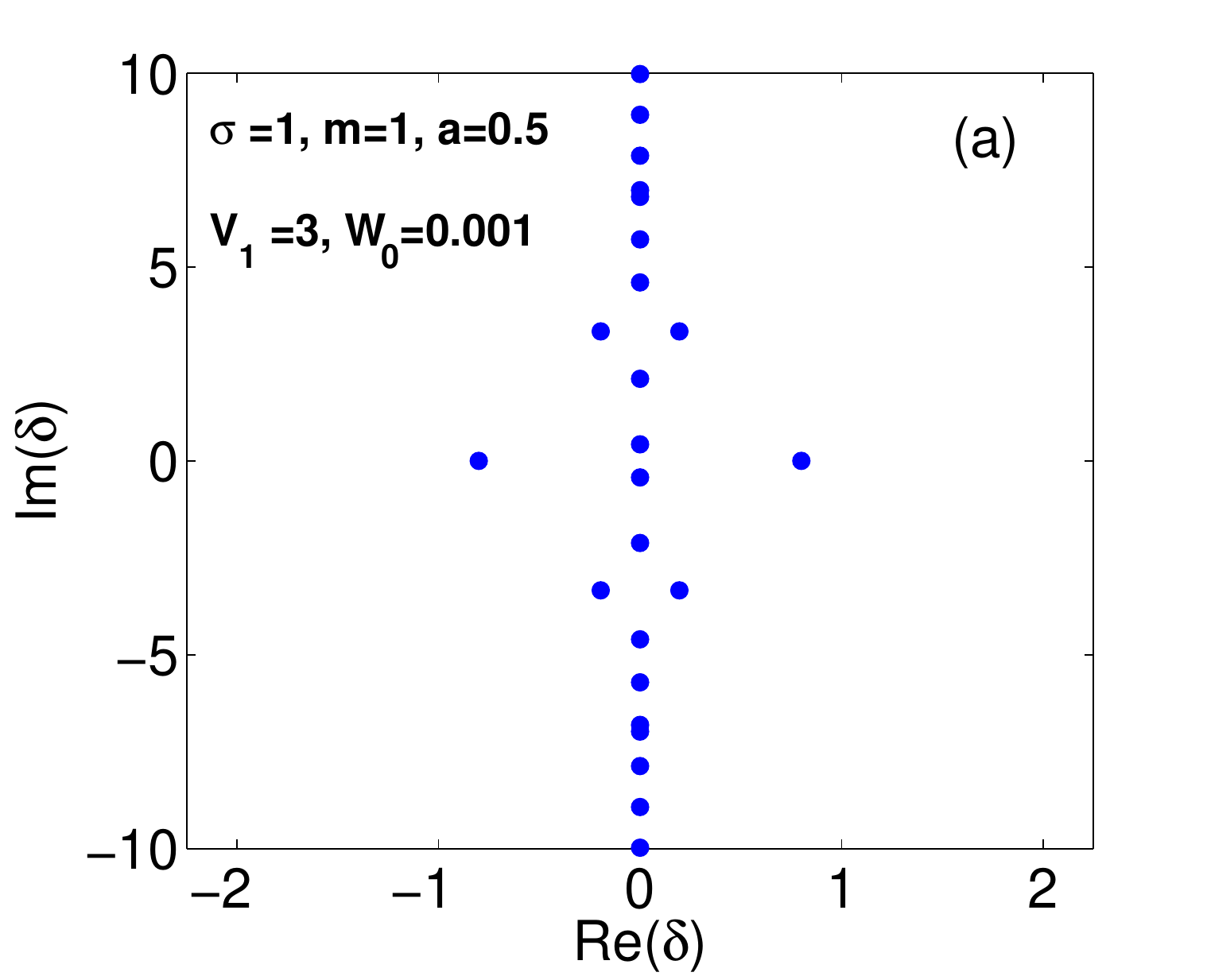} \includegraphics[width=4.3cm,height=4.75cm]{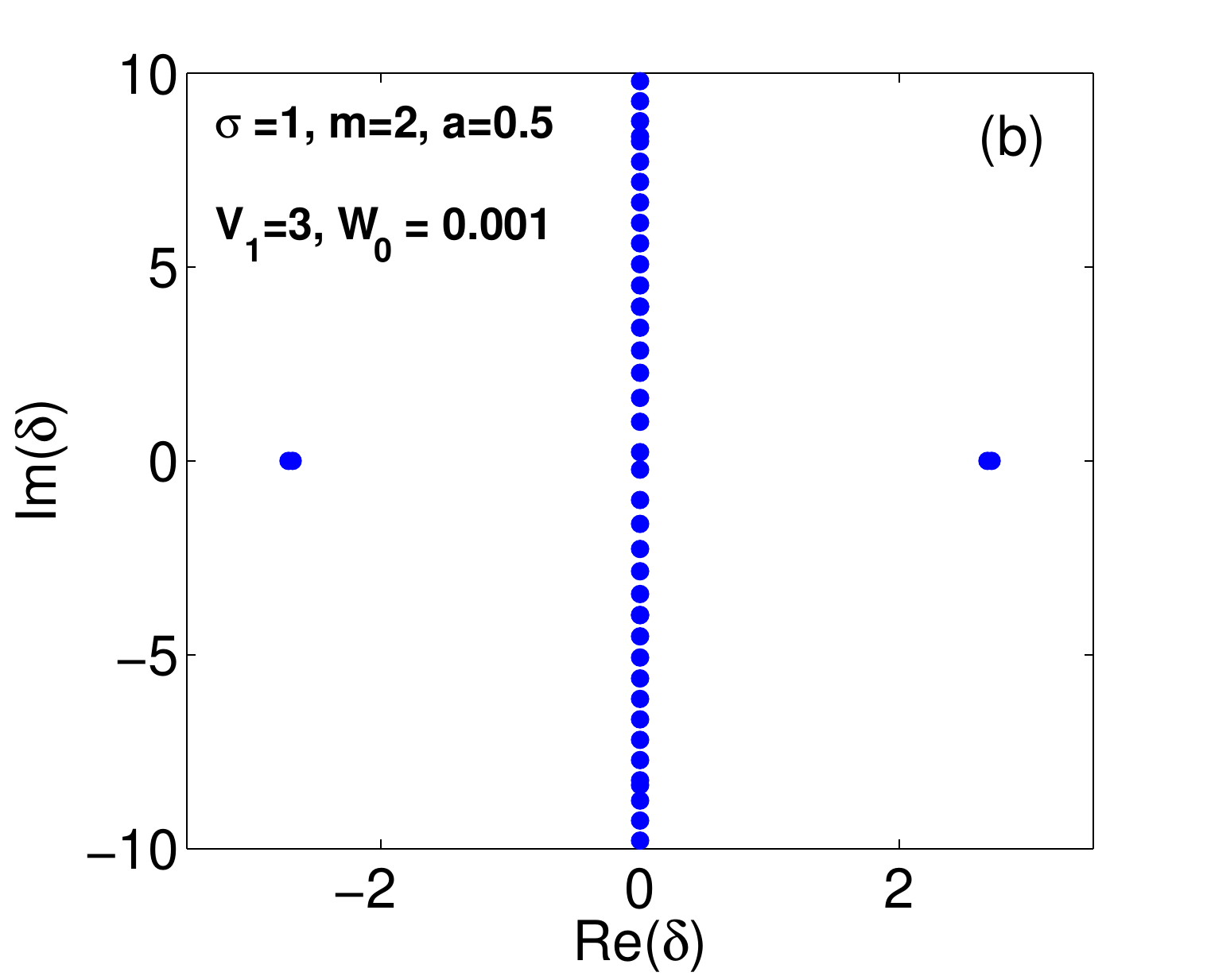}
 \includegraphics[width=4.3cm,height=4.75cm]{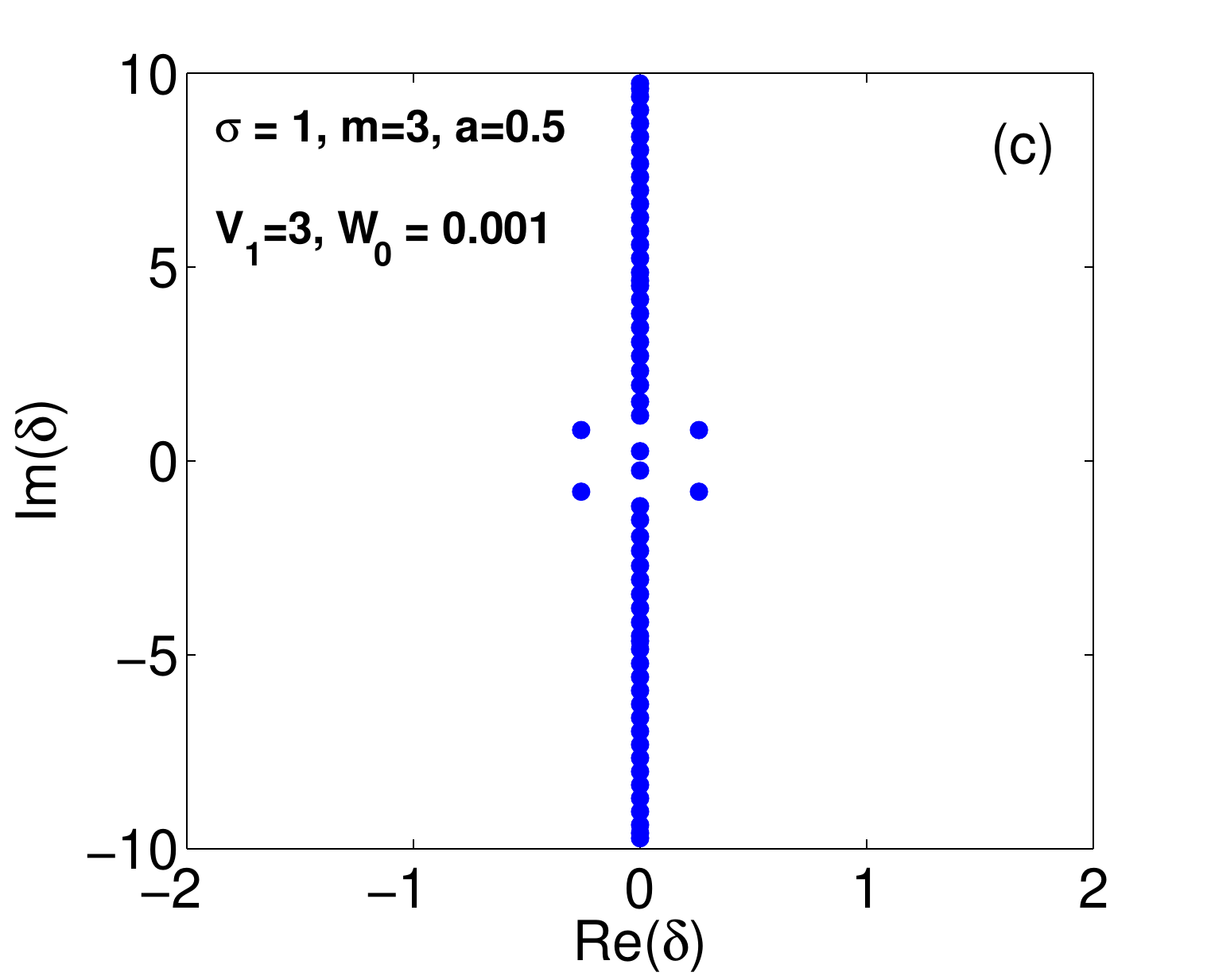}
 \caption{(Color online) Plot of the linear stability spectra in case of self-focusing nonlinearity ($\sigma=1$) for (a) $m=1$, (b) $m=2$ and (c) $m=3$ respectively. Other parameters are chosen as  $a=0.5$, $V_1 =3$ and $W_0 =0.001$.}
 \label{fig3}
 \end{figure}

 For the case of self-focusing nonlinearity, we find that the corresponding linear stability spectra always possess $\delta$ with non-zero positive real part. This suggests that the exact solutions are always unstable regardless of the order of the nonlinearity. The results of our numerical computation for, in particular, $m=1,2,3$ are shown in figure \ref{fig3}. From the figure it is clear that the solutions are unstable even for very very small value of $W_0$ (=0.001). 
\FloatBarrier

\section{Exact Gaussian solitons in 2D PT-symmetric potentials}
Here we investigate the localized modes and their stability supported by a $2D$ PT-symmetric complex potential $V_{PT}(x,y)$ whose real and imaginary parts are given by 
\begin{equation}
\begin{array}{ll}
V(x,y) = V_0 (x^2 + y^2) - V_1 e^{-2 a^2(x^2+y^2)} + V_2 (e^{-2 a^2 x^2} + e^{-2 a^2 y^2})\\
W(x,y) = W_0 (x e^{-a^2 x^2} + y e^{-a^2 y^2}),
\end{array}\label{e13}
\end{equation} 
respectively. Clearly,  $V(-x,-y) = V(x,y)$ and $W(-x,-y) = - W(x,y)$. The stationary solution of the following generalized NLSE in (2+1) dimensions  
\begin{equation}
 i \frac{\partial \Psi}{\partial z} + \nabla^2 \Psi + [V(x,y) + i W(x,y)] \Psi + |\Psi|^{2m} \Psi = 0, \label{e10}
\end{equation}
where $\nabla^2 \equiv \frac{\partial^2 }{\partial x^2} + \frac{\partial^2 }{\partial y^2}$, can be assumed 
in the form
\begin{equation}
 \Psi(x,y,z) = \phi(x,y) ~ e^{i \beta z + i \theta(x,y)}.
\end{equation}
Here the real valued functions phase $\theta(x,y)$, and the soliton $\phi(x,y)$ satisfy the following differential
equations
\begin{equation}\begin{array}{ll}
 \nabla^2 \phi - |\nabla \theta|^2 \phi + V(x,y) \phi + \sigma \phi^{2m+1} = \beta \phi, \\
 \phi\nabla^2 \theta + 2 \nabla \theta . \nabla \phi + W(x,y)\phi  = 0,
\end{array}\label{e22}
\end{equation}
respectively. For the potential (\ref{e13}), the above equation (\ref{e22}) possesses the closed form localized solutions [that satisfy $\phi \rightarrow 0$ as 
$(x,y) \rightarrow \pm \infty$] 
\begin{equation}
 \phi(x,y) =  \left|\frac{V_1}{\sigma}\right|^{\frac{1}{2m}} ~ e^{- \frac{a^2(x^2+y^2)}{m}},
\end{equation}
and the phase is given by
\begin{equation}
\theta(x,y)  =  \frac{m W_0 \sqrt{\pi}}{4 a^3 (m+2)} [\ef(a x)+\ef(a y)],
\end{equation}
where $V_0 = -\frac{4 a^4}{m^2}, V_2 = \frac{m^2 W_0^2}{4a^4(m+2)^2}, \beta = -\frac{4a^4}{m}$. 
\begin{figure}
\includegraphics[width=4.35cm,height=5cm]{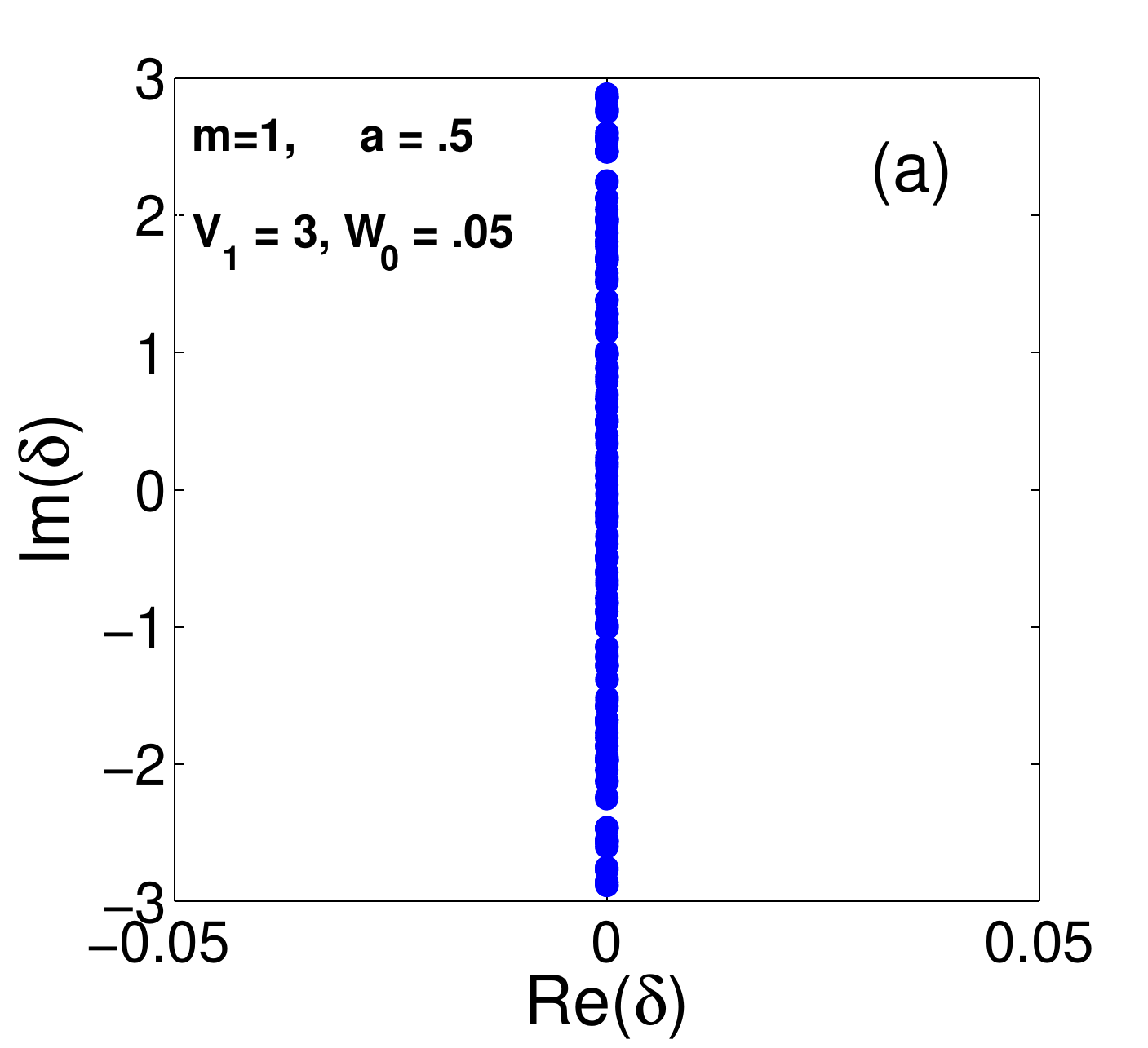}~ \includegraphics[width=5.25cm,height=5cm]{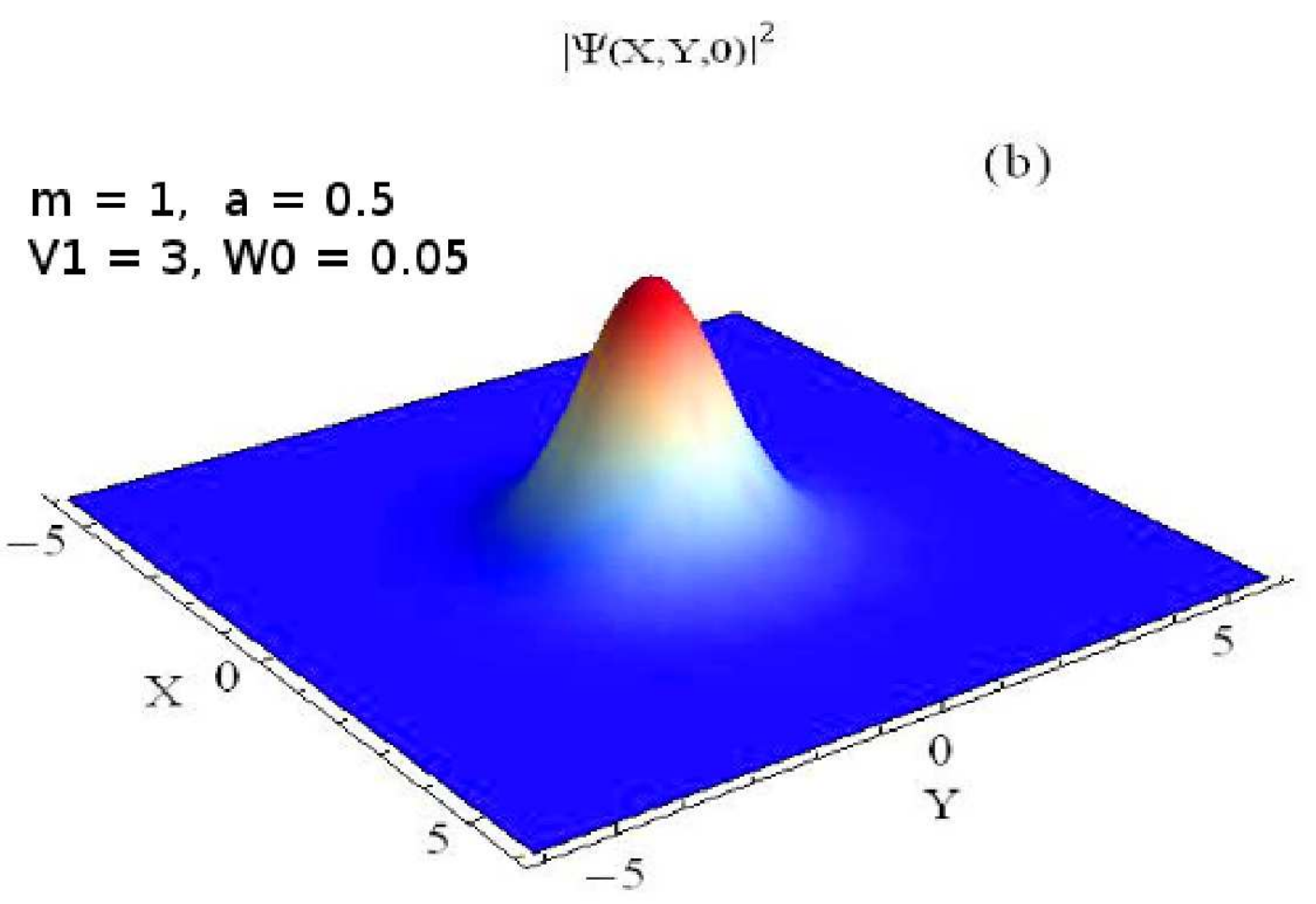}
~\includegraphics[width=5.25cm,height=5cm]{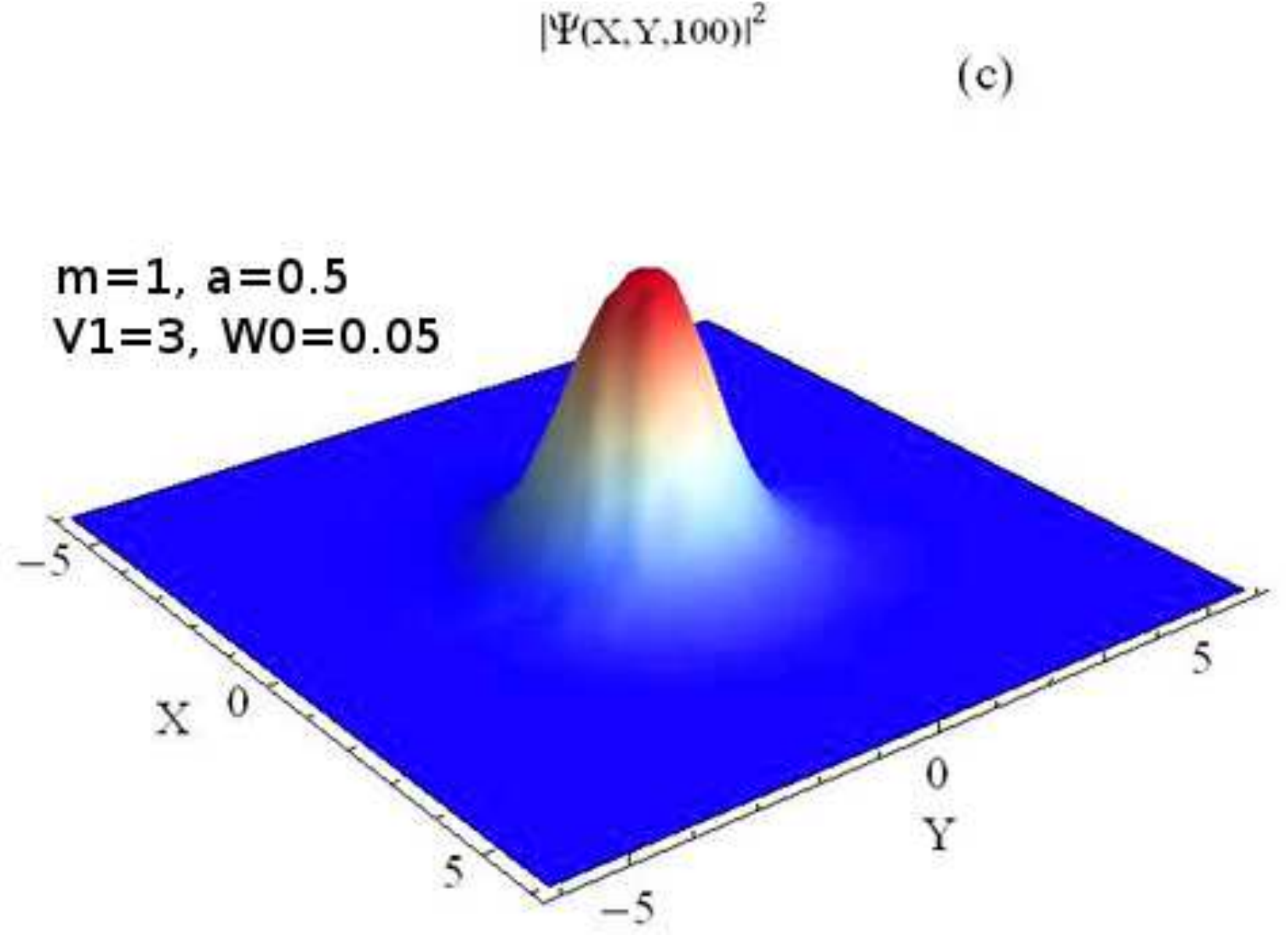} \\
\vspace{.75 cm}
\includegraphics[width=4.2cm,height=5.2cm]{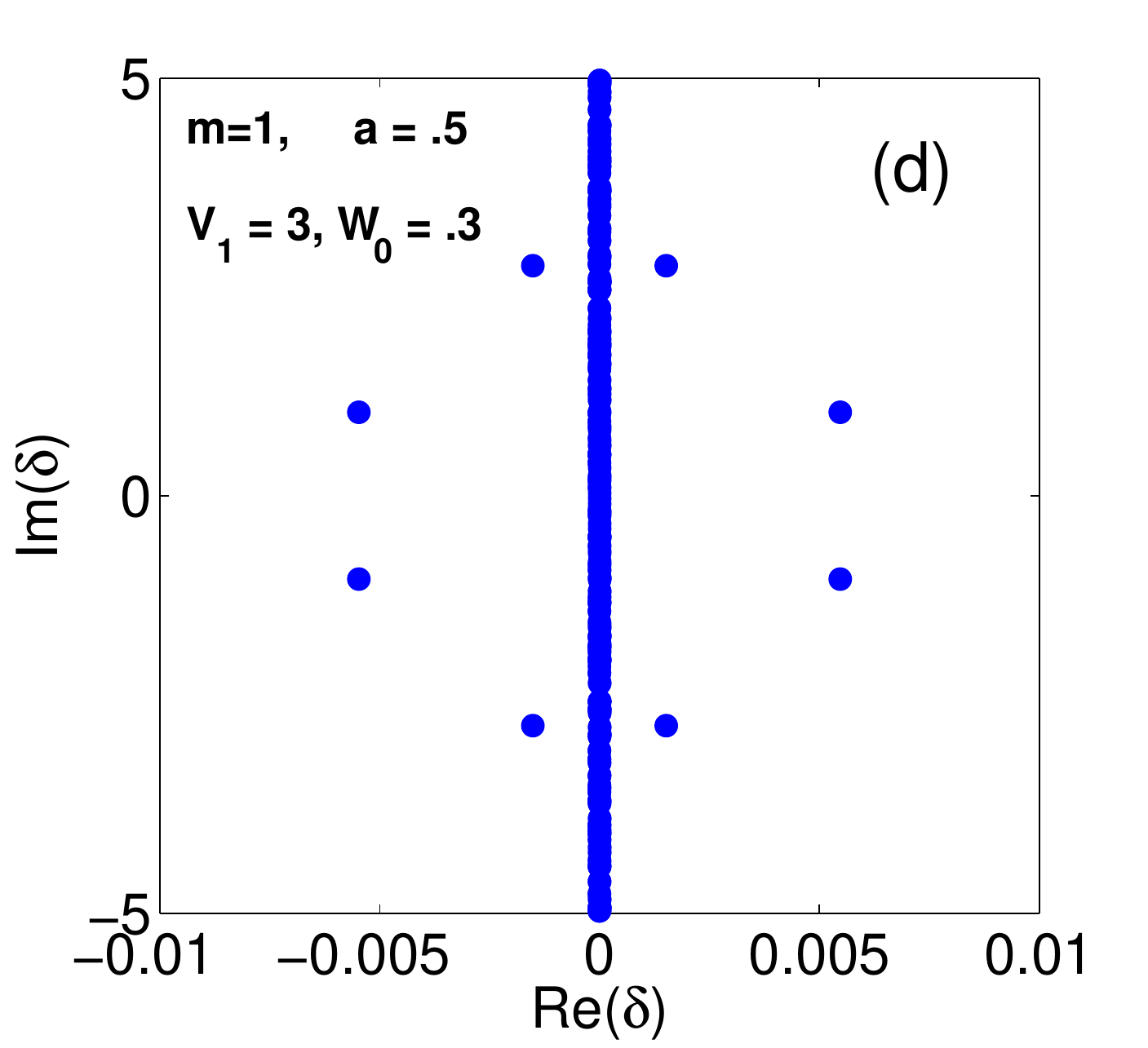}
~\includegraphics[width=5.3cm,height=4.8cm]{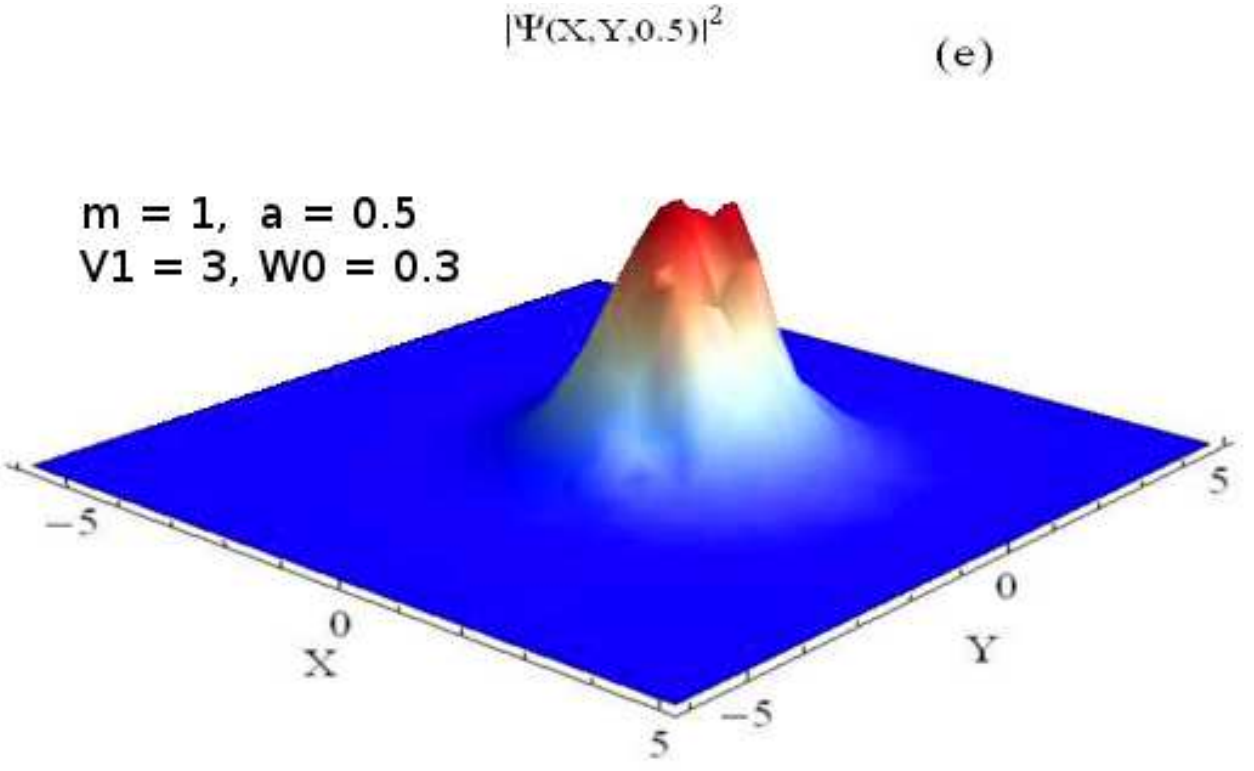} ~ \includegraphics[width=5.5cm,height=5.1cm]{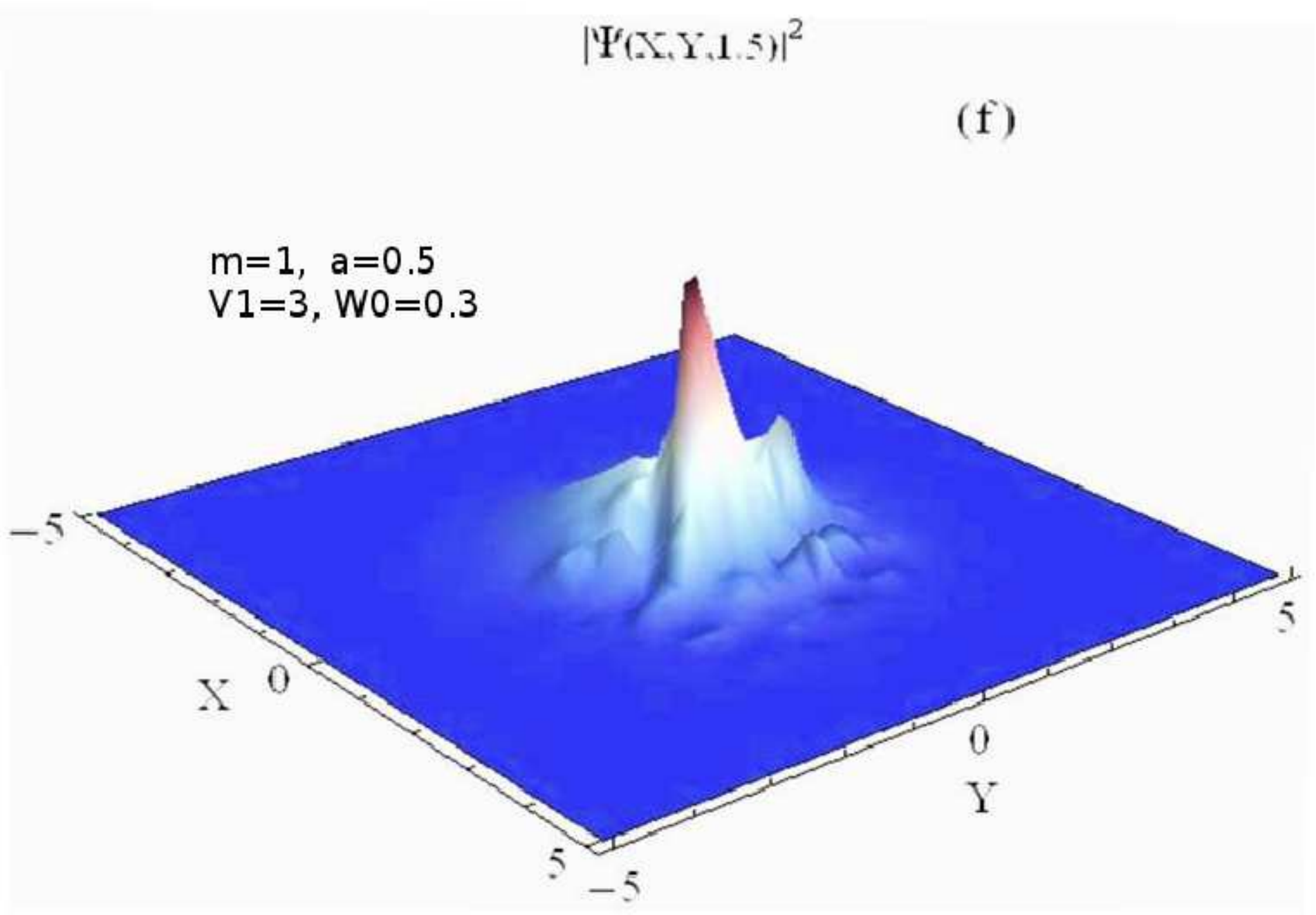}
\caption{(Color online) (a) and (d) Plots of the linear stability eigenvalue spectra below and above the threshold $W_0 = 0.1$, respectively; (b) and (c) plots of the stable intensity evolution at $z=0$ and $z=100$, respectively; (e) and (f) plots of the unstable intensity evolution at $z=0.5$ and $z=1.5$, respectively. In all these cases we have considered $\sigma = -1$.}
\label{fig2}       
\end{figure}
The transverse energy-flow density for these 2D solutions is calculated as $\vec{S} = \left|\frac{V_1}{\sigma}\right|^{1/m} \frac{m W_0 }{2 a^2 (m+2)} \left(e^{-\frac{m+2}{m} a^2 x^2},e^{-\frac{m+2}{m} a^2 y^2}\right)$. Due to $W_0>0$, $\vec{S}$ is everywhere positive. This implies that the energy exchange of the light field across the beam is from gain toward loss regions. 

The linear stability of the 2D localized modes, obtained here, is studied by considering the 2D generalization of the eigenvalue problem given in (\ref{e12}). For numerical computation, we assumed $m=1, \sigma =~-1, a=0.5,$ and  $V_1=3$. In this case, we find that that localized modes are stable for all values of $W_0$ lying below the threshold $W_0 \sim 0.1$. In figures \ref{fig2}(a) and \ref{fig2}(d), we have shown the numerically computed eigenvalue spectra for two values of $W_0$ one is below and another is above the threshold value, respectively. To obtain the corresponding intensity evaluation, we have performed the direct numerical simulation of equation (\ref{e10}) by taking the initial profile as $\Psi(x,y,0) = \phi(x,y) e^{i \theta(x,y)}$. The results are shown in figures \ref{fig2}(b) and \ref{fig2}(c) for stable propagation, and in \ref{fig2}(e) and \ref{fig2}(f) for unstable propagation. Like 1D case, the 2D solitons are also always unstable for self-focusing nonlinearity.

\section{Conclusions}
In conclusion, we have addressed a novel family of exact localized solutions supported by a PT-symmetric complex potential in the presence of higher order nonlinearities including the Kerr one. For self-defocusing nonlinearity, these solutions are shown to be stable for a wide range of the potential parameters in both (1+1) and (2+1) dimensions. It is shown that there exists a threshold value of $W_0$ below which the corresponding solutions propagate stably and above this value, they become unstable. In the former case, the gain (loss) is small enough, compared to the fixed real part of the potential, to suppress the collapse of the solitonic structure caused by the diffraction. In case of self-focusing nonlinearity, we find that the corresponding solutions are always unstable. It is also shown that the nature of the transverse power-flow density, associated with these localized modes, is always positive, implying the energy flow from gain toward the loss. It would be interesting to examine other possible solitonic solutions of this model by means of numerical techniques and to study their stability.

Finally, we would like to comment on the possible realization of the model considered here. In general for practical application of an optical beam propagation, it is desirable to consider a periodic structure. However, the real part of the potential considered here is a single or double well (depending on the parameter values) with a confined imaginary part which vanishes at large distance. Therefore the potential and corresponding exact solution may be used to model a single nonlinear PT cell \cite{Mu+08} or a PT-symmetric Bose-Einstein condensate as suggested in \cite{CG12}.

\newpage
\begin{acknowledgements}
The author thanks Prof. Rajkumar Roychoudhury for discussions. He also acknowledges the postdoctoral grant
from the Belgian Federal Science Policy Office co-funded by the Marie-Curie
Actions (FP7) from the European Commission. 
\end{acknowledgements}

\end{document}